\documentclass[3p,onecolumn,times]{article}
\usepackage{color}
\usepackage[figuresright]{rotating}

\definecolor{vertmat}{rgb}{0,0.9,0}

\usepackage{times}
\usepackage{enumerate}
\usepackage[T1]{fontenc}
\usepackage{url}
\usepackage{natbib}

\usepackage{gensymb}
\usepackage{amsfonts,amssymb,amsmath}
\usepackage{bbm,bm}
\usepackage{graphicx}
\usepackage{mathptmx} 

\newcommand{\bef}{\begin{figure}}
\newcommand{\eef}{\end{figure}}
\newcommand{\bee}{\begin{enumerate}[i]}
\newcommand{\eee}{\end{enumerate}}

\newcommand{\bei}{\begin{itemize}}
\newcommand{\eei}{\end{itemize}}

\newcommand\moy[1]{\mu\left[#1\right]}
\newcommand\pT[2]{\pi^{(T)}_{#2}\left[#1\right]}

\newcommand{\ed}{\end{document}}
\newcommand{\beq}{\begin{equation}}
\newcommand{\eeq}{\end{equation}}
\newcommand{\nid}{\noindent}
\newcommand{\ben}{\begin{enumerate}}
\newcommand{\een}{\end{enumerate}}
\newcommand{\bit}{\begin{itemize}}
\newcommand{\eit}{\end{itemize}}
\newcommand{\baR}{\begin{array}}
\newcommand{\eaR}{\end{array}}

\newcommand{\ssu}[1]{\subsection{#1}}

\newcommand{\bea}{\begin{eqnarray}}
\newcommand{\eea}{\end{eqnarray}}

\newcommand{\pare}[1]{\left(\, #1 \, \right)}
\newcommand{\brac}[1]{\left[\, #1 \, \right]}
\newcommand{\Set}[1]{\left\{\, #1 \, \right\}}

\newtheorem{definition}{Definition}
\newcommand{\bdf}{\begin{definition}}
\newcommand{\edf}{\end{definition}}

\newtheorem{theorem}{Theorem}
\newcommand{\bth}{\begin{theorem}}
\newcommand{\enth}{\end{theorem}}

\newtheorem{proposition}{Proposition}
\newcommand{\bp}{\begin{proposition}}
\newcommand{\ep}{\end{proposition}}
\newtheorem{corrollary}{Corrollary}
\newcommand{\bcor}{\begin{corrollary}}
\newcommand{\ecor}{\end{corrollary}}
\newtheorem{lemma}{Lemma}
\newcommand{\blem}{\begin{lemma}}
\newcommand{\elem}{\end{lemma}}
\newtheorem{conjecture}{Conjecture}
\newcommand{\bconj}{\begin{conjecture}}
\newcommand{\econj}{\end{conjecture}}

\newcommand\bloc[2]{\left[\omega\right]_{#1}^{#2}}

\newcommand\cL{{\cal L}}

\newcommand\mpg{\mu}

\newcommand\scal[2]{\langle #1 \mid #2  \rangle}

\newcommand\rpfc[1]{|b_{#1}\rangle}

\newcommand\rpf{|b\rangle}
\newcommand\rpfw{\rpfc{w}}

\newcommand\lpf{\langle b|}
\newcommand\lpfw{\langle b_w|}
\newcommand\RPF{\cL(\psi)}

\newcommand\RPFww{\cL_{w' , w}(\psi)}
\newcommand\RPFnww{\cL^n_{w' , w}(\psi)}
\newcommand\spsi{s(\psi)}

\newcommand{\ands}{, }

\graphicspath{{sty-fig/}}

\begin{document}

\newcommand{\Olivier}[1]{\textcolor{red}{[#1]}}







\title{Gibbs distribution analysis of temporal correlations structure 
in retina ganglion cells }


\author{ J. C. Vasquez \thanks{NeuroMathComp team (INRIA, ENS Paris, UNSA LJAD), Sophia Antipolis, France. \newline \indent INRIA, 2004 Route des Lucioles, 06902 Sophia-Antipolis, France. \newline \indent   email: Juan-Carlos.Vasquez@sophia.inria.fr}, O. Marre \thanks{Department of Molecular Biology and Princeton Neuroscience Institute, Princeton University, USA and NeuroMathComp team (INRIA, ENS Paris, UNSA LJAD) } , A.G. Palacios \thanks{ Centro Interdisciplinario de Neurociencia de Valparaiso, Universidad de Valparaiso, Chile} ,  M.J. Berry II  \thanks{Department of Molecular Biology and Princeton Neuroscience Institute, Princeton University, USA}  \, and B. Cessac $^\ast$ }

\maketitle

\begin{abstract}
We present a method to estimate Gibbs distributions with \textit{spatio-temporal} constraints
on spike trains statistics.  
We apply this method to spike trains recorded from ganglion cells of the salamander retina, in response to natural movies. 
Our analysis, restricted to a few neurons, performs more accurately than pairwise synchronization models (Ising) or the 1-time step Markov models (\cite{marre-boustani-etal:09}) to describe the statistics of spatio-temporal spike patterns and emphasizes the role of higher order spatio-temporal interactions.\\

\end{abstract}




\section{Introduction}
\label{Sec:intro}

	Modern advances in neurophysiology techniques, such as two-photons imaging of calcium signals or micro-electrode arrays electro-physiology, have made it possible to observe simultaneously the activity of assemblies of neurons, \cite{stevenson-kording:11}. Such experimental recordings provide a great opportunity to unravel the underlying interactions of neural assemblies. The analysis of multi-cells spike-patterns constitutes an alternative to descriptive statistics (e.g cross-correlograms or joint peri-stimulus time histograms) which become hard to interpret for large groups of cells, \cite{brown-etal:04,kass-etal:05}. Earlier multi-cells approaches, e.g., \cite{Abeles88}, focus on synchronization patterns. Using algorithms detecting the most frequent instantaneous patterns in a data set, and calculating their expected probability,
these approaches aim at testing whether those patterns were produced by chance, \cite{Grun2002}.  This methodology relies however on a largely controversial assumption, namely Poisson-statistics,  \cite{pouzat-chaffiol:09,schneidman-berry-etal:06}.\\

A second type of approach has become popular in neuroscience after works of \cite{schneidman-berry-etal:06,shlens-field-etal:06}. They used a maximum entropy approach to model spike trains statistics by the Gibbs distribution of the Ising model. The parameters of this distribution are determined from the mean firing rate of each neuron and their pairwise synchronizations.
These works have shown that for a small group of cells (10-40 retinal ganglion cells) the Ising model  describes most ($\sim 80-90\%$) of the statistics of the \textit{instantaneous} patterns, and performs much better than a non-homogeneous Poisson model.

However, several papers have pointed out the importance of temporal patterns of activity at the network level , \cite{abeles-etal:93,lindsey-morris-etal:97,villa-tetko-etal:99,segev-baruchi-etal:04}. Recently, \cite{tang-etal:08,ohiorhenuan-etal:10}, have shown the insufficiency of the Ising model to predict the temporal statistics of the neural multi-cells activity. Therefore, some authors, \cite{marre-boustani-etal:09,amari:10,roudi-hertz:10}, have attempted to define time-dependent Gibbs distributions on the basis of a Markovian approach (1-step time pairwise correlations). The application of such extended model in \cite{marre-boustani-etal:09} increased the accuracy of the statistical characterization of data with the estimated distributions.

In this paper we propose an extension of the maximal entropy approach to general spatio-temporal correlations, based on the transfer-matrix method in statistical physics, \cite{georgii:88} (section 2). We describe a numerical method to perform the estimation of the Gibbs distribution parameters from empirical data (section 3). 
We apply this method to the analysis of spike trains recorded from ganglion cells  using multi-electrodes devices in the salamander retina (section 4). We analyse retinal spike trains taking into account spatial patterns of two and three neurons with triplets and quadruplets terms, and temporal terms up to $4$ time steps.  Our analysis emphasizes the role of higher order spatio-temporal interactions.
Section 5 contains the discussion and conclusions.

\section{Theoretical framework}
\label{stheory}
\ssu{Spike trains and Raster Plots}
Let $N$ be the number of neurons and
denote $i=1, \dots, N$  the neuron index. 
Assume that we have discretised  time in  steps of size $\Delta$. Without loss of generality (change of time units) we may set $\Delta=1$.  This provides
a time discretisation labelled with an integer index $n$. We define a binary variable $\omega_i(n) \in \{0,1\}$, which is '$1$' if neuron $i$ has emitted a spike in the $n$-th time interval and is zero otherwise. We use the notation $\omega$ to differentiate our binary variables $\in \{0,1\}$ 
to the notation $\sigma$ or $S$ traditionally used for ``spins'' variables $\in \{-1,1\}$. 
The \textit{spiking pattern} of the neural network at time $n$ is the vector 
$\omega(n)=\left(\omega_i(n)\right)_{i=1}^N$.
We denote  $\bloc{m}{n}$ the ordered sequence
or \textit{spike block } $\omega(m) \dots \omega(n)$, $m \leq n$. 
In practice, from recordings and after applying spike sorting algorithms, one obtains a sequence of spiking patterns called a \textit{raster plot}. In our notations a raster plot is thus a spike block $\bloc{0}{T}$ where $T$ is the total length of
the spike time sequences, measured in $\Delta$ time-units.\\

\ssu{Observables and monomials}
We call \textit{observable} a function $\phi$ which associates to a raster a real number. Although the method developed here holds for general functions, we focus on observables called \textit{monomials}.
These are functions of the form  $\phi(\omega) \,= \, \omega_{i_1}(n_1) \, \omega_{i_2}(n_2) \dots \, \omega_{i_m}(n_m)$ 
which is equal to $1$ if and only if neuron $i_1$ fires at time $n_1$, $\dots$, neuron $i_m$ fires at time $i_m$ in the raster $\omega$. Thus monomials attribute the value '$1$' to characteristic spike events. We use the convention that $n_1 \leq n_2 \leq \dots \leq n_m$. Then,
the \textit{range} of a monomial is 
$n_m - n_1 +1$. 

A typical monomial is  $\phi(\omega) \, = \, \omega_{i}(0)$ which is equal to '$1$' if neuron $i$ spikes at time $0$ in the raster $\omega$ and is '$0$' otherwise. This a function
of a single event, of range $1$. Likewise  $\phi(\omega) \, = \, \omega_{i}(0) \, \omega_{j}(0)$ is '$1$' if and only if neuron $i$ and $j$ fire synchronously at time $0$ in the raster $\omega$. This is a function of pairwise event, of range $1$ too. As a last example, $\phi(\omega) \, = \, \omega_{1}(0) \, \omega_{2}(1)\, \omega_{3}(2)\, \omega_{4}(5) $
is a function of a quadruplet of spikes, of range $6$.\\

\ssu{Hidden probability}
Collective neuron dynamics, submitted to noise, produces spike trains with randomness, although
some statistical regularity can be observed. The spike trains statistics are assumed
to be characterized by an hidden probability  $\mu^{h}$ giving the probability of \textit{spatio-temporal} spike patterns.
A current goal in experimental analysis of spike trains is to approximate $\mu^{h}$ from data. A \textit{model} is a probability distribution $\mu$ which approaches $\mu^{h}$.
We give a precise meaning of approaching a probability by another one below.
Typically, $\mu$ must predict the probability of spike blocks occurrence with a good accuracy. 

Given a model $\mu$ we note $\moy{\phi}$ the average of an observable $\phi$ with respect to $\mu$. For example the average
value of $\phi(\omega) = \omega_i(n)$ is given by 
$\moy{\omega_i(n)} \,= \, \sum_{\omega_i(n)}
\omega_i(n) \, \moy{\omega_i(n)}$ where the sum holds on all possible values of $\omega_i(n)$ ($0$ or $1$). Thus, finally 
$\moy{\phi} = \moy{\omega_i(n)=1}$ is nothing but the probability of firing of neuron $i$ at time $n$, predicted by the model $\mu$.
Likewise, the average value of $\omega_{i_1}(n) \, \omega_{i_2}(n)$ is the predicted probability that neuron $i_1$  and $i_2$
fire at the same time $n$: this is a measure of pairwise synchronization. More generally, for the monomial $\phi=\omega_{i_1}(n_1) \, \omega_{i_2}(n_2) \dots \, \omega_{i_m}(n_m)$,
$\moy{\phi}$ is the predicted probability of occurrence of the event "neuron $i_1$ fires at time $n_1$, $\dots$, neuron $i_m$ fires at time $i_m$".

We assume here, as in most papers dealing with spike train statistics, that hidden statistics are stationary so that the average value of functions is time-translation invariant. As a consequence we consider time-translation invariant models (e.g., $\moy{\omega_i(n)=1}$ is independent on $n$).  \\

\ssu{Time-average}
Given an experimental raster $\omega$ of duration $T$, and an observable $\phi$
we note $\pT{\phi}{\omega}$ the time-average of $\phi$. For example, when $\phi(\omega) \, = \, \omega_i(n)$, $ \pT{\phi}{\omega} \,=\,  \frac{1}{T} \, \sum_{n=0}^{T-1} \omega_i(n)$ provides an estimation of the firing rate of neuron $i$ (it is independent of time from the stationarity assumption). If $\phi$ is a monomial $\omega_{i_1}(n_1) \dots \omega_{i_m}(n_m)$, $1 \leq n_1 \leq n_2 \leq n_m < T$ then 
$\pT{\phi}{\omega} \,=\,  \frac{1}{T-n_m} \, \sum_{n=0}^{T-n_m} 
\omega_{i_1}(n_1+n) \dots \omega_{i_m}(n_m+n)$, and so on. We use the cumbersome notation $\pT{\phi}{\omega}$ to remind that such time averages
are \emph{random variables}. They fluctuate from one raster to another and the amplitude of those fluctuations depend on $T$.
We assume ergodicity which is a common hypothesis in this field.  Then, for any observable $\phi$,  $\pT{\phi}{\omega} \to \mu^{h}\left[\phi\right]$ as $T \to +\infty$, where the limit is independent of the raster $\omega$.\\

\ssu{Gibbs distribution}
Fix a set of observables $\{\phi_l\}_{l=1}^{L} $ whose time average $\pT{\phi_l}{\omega}$ has been measured and is equal to $C_l$.
To match those empirical statistics, the model $\mu$ has to satisfy: 
\beq \label{Constraints}
\moy{\phi_l} \,=\, \pT{\phi_l}{\omega} \,=\, C_l, \quad l=1, \dots, L.
\eeq
This is a minimal, but insufficient requirement, since one can construct infinitely
many probability distributions satisfying the constraints (\ref{Constraints}). 

However, with the additional requirement that the model has to ``Maximize the statistical entropy under the constraints (\ref{Constraints})'', a unique model is selected. This is the maximal entropy principle,  \cite{jaynes:57} that amounts to solving, i.e. find the maximum of: 
a \textit{variational principle}:
\beq\label{varprinc}
P(\psi)=\sup_{\nu \in m^{(inv)}} \pare{h\left[\nu\right]+
\nu\left[\psi\right]}.
\eeq
The term $\psi$ defined by:
\beq\label{psi}
\psi=\sum_{l=1}^L \lambda_l \phi_l,
\eeq
is called a  \textit{potential}. The $\lambda_l$ are free parameters (Lagrange multipliers). $\psi$ is thus a linear combination of the observables defining the constraints (\ref{Constraints}). The supremum in (\ref{varprinc}) is taken over $m^{(inv)}$, the set of time-translation invariant (stationary) probabilities on the set of rasters for $N$ neurons.  $h$ is the  
entropy rate, see \cite{ruelle:69,ruelle:78,keller:98,chazottes-keller:09} for the general definition.

 A probability $\mpg$ which realizes the supremum (\ref{varprinc}), i.e.,
\beq\label{pres}
P(\psi)=h\left[\mpg\right]+\mpg\left[\psi\right],
\eeq
is called a \textit{Gibbs distribution}. 
This name has its roots in statistical physics and we discuss this connection in the next paragraph.
The term $P(\psi)$, called the \textit{topological pressure} in this context is the formal
analog of a thermodynamic potential (free energy density). It is a generating function for the cumulants
of $\psi$. In particular;
\beq\label{moy_phil}
\frac{\partial P(\psi)}{\partial \lambda_l}  \, = \, \moy{\phi_l}.
\eeq

Let us summarize what we have just obtained. To a set of experimental constraints, associated with a set of observables $\{\phi_l\}_{l=1}^{L} $, one associates a probability distribution $\mpg$, called a Gibbs distribution, parametrized by the potential (\ref{psi}), a linear combination of $\phi_l$'s. Now, comparing
equations (\ref{Constraints}) and (\ref{moy_phil}), one sees that the free parameters $\lambda_l$ can be adjusted so that the Gibbs distribution $\mpg$ matches the constraints (\ref{Constraints}). We will explain how this computation can be done in section \ref{Sestimation}. It turns out that $P(\psi)$ is a convex function. Therefore, there is a unique set of $\lambda_l$ so that $\mpg$ matches the constraints (\ref{Constraints}).
Hence the maximal entropy principle provides 
a unique statistical model matching the experimental constraints (\ref{Constraints}).

Note that $\mpg$ depends on $\psi$, thus (i) on the choice of observables; (ii) on the parameters $\lambda_l$. However, we drop this dependence in the notation to ease legibility.\\
 
\ssu{A remark. Links with previous approaches}
The maximal entropy principle is commonly used in statistical physics and has been applied by several authors for spike trains
analysis, \cite{schneidman-berry-etal:06,tkacik-schneidman-etal:09,tkacik-prentice-etal:10,schaub-schultz:10,ganmor-segev-etal:11a,ganmor-segev-etal:11b}. Here we would like to insist on the main difference between our approach and the one of these authors.

In those references, constraints correspond to
\textit{simultaneous} spike events (monomials of the form $\omega_{i_1}(n) \dots \omega_{i_m}(n)$) corresponding to \textit{spatial} patterns. On the opposite, our observables
$\phi_l$ correspond to  spatio-temporal events
so that $\psi$ depends on the raster plot over a (finite)
time horizon $R$, i.e. $\psi(\omega) \equiv \psi(\bloc{0}{R-1})$.
We speak of ``range-$R$ potentials''. Thus, our method imposes constraints on general \textit{spatio-temporal} events instead of focusing on spatial constraints.

The difference is not anecdotic. Imposing spatio-temporal constraints
amounts to considering a statistical model in which
the probability of a spiking pattern depends on the past history: the system has a memory and its actual state depends on its past via a set of causal spatio temporal relations. Typically, this is described by a Markovian process,
(although non Markovian dynamics also occur in neural networks models, \cite{kravchuk-vidybida:10,cessac:11a,cessac:11b}). 
The Markovian case has been considered by several authors in the field of spike statistics analysis, but with one time step memory only, and under assumptions such as detailed balance,
\cite{marre-boustani-etal:09} or conditional independence between neurons, see eq. (1) in \cite{roudi-hertz:11}. 
 
The method introduced here does not use these assumptions and allows us to consider, on a theoretical ground, general spatio-temporal constraints. It is based on a mathematical object called, in statistical physics, "transfer matrix" \cite{georgii:88} and in ergodic theory "Ruelle-Perron-Frobenius operator", \cite{bowen:75,ruelle:78,meyer:80}. Although this method extends to non-Markovian dynamics, \cite{cessac:11a,cessac:11b}, in the present paper, we restrict to finite memory. In this restricted case, this method has its roots in matrix representation of Markov chains and Perron-Frobenius theorem, \cite{gantmacher:66,seneta:06}. So this method is well known but, to our knowledge, it is the first time that it is applied to the analysis of spike trains. 

Since we focus on Markovian dynamics here, Gibbs distribution could also be introduced in this setting, see \cite{cessac-palacios:11} for a didactic presentation in the realm of spike train analysis. However, the advantage of the presentation adopted here is its compactness compatible with the limited allowed space of the paper.

\section{Estimation of Gibbs Distributions}\label{Sestimation}

Let us now show how $P(\psi)$ and $\mu$, the main objects of our approach, can be computed.

\ssu{The  transition matrix}
The range $R$ of the potential $\psi$ is the maximum of the ranges of monomials defining $\psi$.
If $R=1$ the potential depends only on simultaneous events and corresponds to considering a memory-less process as a model. On the opposite, if $R>1$ the potential accounts for spatio-temporal events and corresponds to taking into account memory and time-causality in the model. 
 
We assume here that $R>1$ and come back to the case $R=1$ below.
The starting point is to consider  that a block $\bloc{n}{n+R-1}$
is a transition from a block $\bloc{n}{n+R-2}$, of range $R-1$, to a block
$\bloc{n+1}{n+R}$ of range $R-1$ too. Therefore, the two blocks overlap (the sequence $\bloc{n+1}{n+R-1}$ is common to both blocks).
It is useful to choose a symbolic representation of spike blocks of range $R-1$. Indeed,
there are $M=2^{N(R-1)}$ such possible spike blocks,
requiring, to be represented, $N(R-1)$ symbols ('0''s and '1''s). Instead, we associate to
each block $\bloc{n}{n+R-1}$ an integer:
\beq \label{omtow}
w_n
=\sum_{r=0}^{R-1} \sum_{i=0}^{N-1} 2^{i+Nr}\omega_i(n+r).
\eeq
We write $w_n \sim \bloc{n}{n+R-1}$.
Now, for integer $m,n$ such that $m \leq n$, $n-m \geq R$, 
a spike sequence 
$\bloc{m}{n}=
\omega(m) \, \omega(m+1) \dots \omega(m+R-1) \dots \omega(n)$  can be encoded
as a sequence of integers $w_m,w_{m+1} \dots w_{n-R+1}$. Clearly, this representation introduces
a redundancy since successive blocks $w_n, w_{n+1}$ have a strong overlap. But what we gain is a convenient
matrix representation of the spike trains process.
Note that each symbol $w_n$ belongs to $\Set{0, \dots, 2^{N(R-1)}}$. However, when encoding spike trains by a sequence of such symbols, we cannot have any possible 
succession of symbols $w_n, w_{n+1}$. Indeed, the corresponding blocks must overlap (they must have the sequence $\bloc{n+1}{n+R-1}$ in common). We say that the 
succession $w_n, w_{n+1}$ is \textit{legal} if the corresponding blocks overlap. 

For two integers $w',w \in \Set{0, \dots, 2^{N(R-1)}}$,
we define the \textit{transition matrix} $\RPF$ with entries:
\beq\label{defRPF}
 \RPFww=   
\left\{
\baR{lll}
 e^{\psi_{w'w}}
\quad &\mbox{if} \quad
w', w &\mbox{is legal} \\
0, \quad &\mbox{otherwise}.
\eaR
\right.,
\eeq
where $\psi_{w'w}$ stands for
 $\psi(\bloc{0}{R-1})$. Indeed, for a legal transition $w',w$, fixing $w', w$  is equivalent to fixing the block $\bloc{0}{R-1}$  as well.\\

\textbf{Remark.} $\RPF$ is a huge matrix (with $2^{N(R-1)} \times 2^{N(R-1)}$ symbols). However,
\begin{enumerate}
\item This is a \textit{sparse} matrix. Indeed, on a each row, there are at most $2^N$ non-zero entries.
\item If, instead of considering all possible symbols, one restricts to symbols (blocks) \textit{effectively}
appearing in an experimental raster, the dimension is considerably reduced.
\end{enumerate}

\ssu{The  Perron-Frobenius theorem}
Since, $\RPF$ is a positive  \textit{matrix} it obeys the Perron-Frobenius theorem, \cite{gantmacher:66,seneta:06}. Instead of stating it in its full generality, we give it under the assumption  that the $\RPF$ is  primitive,
i.e. $\exists n>0$, s.t. $\forall w,w'$ $\RPFnww>0$. This assumption holds for Integrate and Fire models with noise  and is likely to hold for more general neural networks models where noise renders dynamics ergodic and mixing.
Then, the Perron-Frobenius theorem states that $\RPF$
has a unique real positive maximal eigenvalue $\spsi$ associated with a right eigenvector $\rpf$ 
and a left eigenvector $\lpf$ such that
$\RPF \rpf=\spsi\rpf$, and $\lpf \RPF =\spsi\lpf$. Those vectors can be chosen such
that the scalar product $\scal{b}{b}=1$.
The remaining part of the spectrum is located in a disk in the complex plane, of radius strictly lower than $\spsi$.

It can be shown, \cite{bowen:75,ruelle:78,keller:98}, that the topological pressure $P(\psi)$
is:
\beq
P(\psi) \,=\, \log \spsi.
\eeq
Moreover, the Gibbs distribution $\mu$ is: 
\beq\label{invmeas}
\mu=\rpf \lpf,
\eeq
i.e. the probability of a spike block $\sim w$ of range $R-1$ is 
$$
\mu(w)=\rpfw \lpfw,
$$
\nid where $ \rpfw$ is the $w$-th component of $\rpf$.
So we have a simple way to compute the topological pressure and the Gibbs distribution by building the transition matrix of the model. Note that we don't have to compute a partition function. 

\subsection{The case $R=1$}

Our method can be applied to this case as well, although other methods for range $1$ potentials have been applied in the literature and are more efficient \cite{schneidman-berry-etal:06,tkacik-schneidman-etal:09}. In our setting, a range-$1$ potential $\psi$ depends on $w'$ only and the matrix $\cL(\psi)$ has
constant non zero coefficients $e^{\psi_{w'}}$ on each raw. Then, it is straightforward to check that this matrix has $N-1$ eigenvalues equal to $0$, while the largest one
is $Z=\sum_{w'} e^{\psi_w'}$, the partition function of a lattice model with potential $\psi$. So $P(\psi)=\log Z$. Likewise the left eigenvector is $\lpf=\left(1, \dots, 1\right)$, while the right eigenvector has entries $\rpfw=e^{\psi_w}$. Thus,
the corresponding probability is $\mu_w=\frac{1}{Z} e^{\psi_w}$, the Gibbs distribution on a lattice, with potential $\psi$.

\ssu{Comparing several Gibbs statistical models}\label{Scomparison}

The choice of a potential (\ref{psi}), i.e. the choice
of a set of observables, fixes a statistical model. Since, there are many choices
of potentials one needs to propose a criterion to compare them. 
  
The Kullback-Leibler (KL) divergence $d_{KL}(\mu,\nu)$
provides some notion of asymmetric ``distance'' between two probabilities,  $\mu$ and $\nu$.
The computation
of $d_{KL}(\mu,\nu)$ is numerically delicate but, in the present context,
the following holds.
For $\nu$ a time-translation invariant probability and $\mpg$ a Gibbs measure with a potential
$\psi$, one has, 
\cite{keller:98,chazottes-keller:09}:  
$$
d_{KL}\left(\nu,\mpg \right) = P(\psi) - \nu\brac{\psi} - h(\nu).
$$

This allows to estimate the divergence of our model $\mu$ to the hidden probability $\mu^h$, providing the exact spike train statistics. The smaller the quantity
$d_{KL}(\mu^h,\mpg)=P(\psi)-\mu^h\brac{\psi} -h(\mu^h)$, the better is the model.
Obviously, since $\mu^h$ is unknown this criterion looks useless. However, 

\begin{enumerate}
\item As stated in the section "Time average", $\mu^h\brac{\psi}$ is well approximated by
$\pT{\psi}{\omega} \, = \, \sum_{l=1}^L \lambda_l
\pT{\phi_l}{\omega}$, where, by definition $\pT{\phi_l}{\omega} \, =\, C_l$. Therefore, $\mu^h\brac{\psi} \sim \sum_{l=1}^L \lambda_l \, C_l$, where $\sim$ means that the right-hand side
approaches the left-hand side as $T \to \infty$.
More precisely, the distance between the two quantities converges to $0$ as $\frac{K_l}{\sqrt{T}}$ \cite{bowen:75,ruelle:78,georgii:88}.

\item The entropy $h(\mu^h)$ is unknown and its estimation by numerical algorithms becomes more and more cumbersome and unreliable as the number of neuron increases, \cite{grassberger:89,schurmann-grassberger:96,gao-kontoyiannis-etal:08}. However, when comparing two statistical models $\mu_1,\mu_2$ with potentials $\psi_1,\psi_2$, for the analyse the \textit{same} data, $h(\mu)$ is a constant since it only depends on data. 
Thus, comparing these two models amounts to comparing $P\left[\psi_1\right]-\pT{\psi_1}{\omega}$ and $P\left[\psi_2\right]-\pT{\psi_2}{\omega}$. 
Introducing
\beq \label{cross-entropy} 
\tilde{h}\left[\psi\right] \, = \, P\left[\psi\right] - \pT{\psi}{\omega} \,=\,  P\left[\psi\right] -  \sum_{l=1}^L \lambda_l \, C_l,
\eeq
(where the $\lambda_l$ depend on the potential via (\ref{moy_phil})), the comparison of two statistical models $\psi_1, \psi_2$, 
i.e. determining if model $\psi_2$ is significantly ``better'' that model $\psi_1$, reduces to the condition:
\begin{equation} \label{comparison} 
\tilde{h}\left[\psi_2\right] \ll \tilde{h}\left[\psi_1\right].
\end{equation}
\end{enumerate}

The advantage of (\ref{cross-entropy}) (sometimes called "cross-entropy") compared to
the KL divergence is that we have removed the entropy, which is subject to huge fluctuations when determined numerically from a finite raster with many neurons. Thus, (\ref{cross-entropy}) is less sensitive to statistical bias.
What we loose, is an \textit{absolute} criterion for
model comparison. We can just say that a model is a better than another one but we cannot say how close 
we are  from the hidden probability. For this latter purpose, we estimate the entropy explicitly  using  
the method proposed by \cite{strong-koberle-etal:98}.
An example is given below, for a small number of neurons.

\ssu{Numerical implementation}
\label{SNum}

Let us now briefly discuss how to numerically estimate
the Gibbs distribution. For details see \cite{vasquez-vieville-etal:10}. The code is available at \url{http://enas.gforge.inria.fr/}.
The algorithmic procedure proposed decomposes in three steps. 

\begin{itemize}

\item\textbf{Choosing a statistical model}, i.e. choosing a guess potential $\psi=\sum_{l=1}^L \lambda_l \phi_l$
or equivalently, a set of observables.

%
%
%
%
%
%

\item\textbf{Computing the time averages $C_l$.}
To compute the time-average we use a data structure of  tree type, with depth $R$ and degree $2^N$, see e.g., \cite{grassberger:89} for a formal introduction. The nodes count the number of occurrences of blocks encountered in the raster. Thus, we do not store explicitly blocks of occurrence zero. Moreover, when comparing the distributions for distinct ranges $R$ we can count in one pass, and in a unique data structure, block  of different ranges. 

\item\textbf{Performing the parametric estimation.}
The parametric estimation aims at finding the $\lambda_l$ minimizing  ~(\ref{cross-entropy}), by calculating the topological pressure. Note that, from
(\ref{cross-entropy}), finding a point where $\tilde{h}$ is extremal is equivalent to solving (\ref{moy_phil}). Additionally, $P(\psi)$ is convex, thus $\tilde{h}$ is convex too as a linear combination of convex functions. Thus, there is a unique minimum corresponding to the solution of (\ref{moy_phil}). 

We start with a random guess for the $\lambda_l$, and then iterate the following steps: 
\begin{enumerate}
\item Build the matrix $\cL(\psi)$ from the values of $\lambda_l$ and equation (\ref{defRPF}).
\item Compute the eigenvectors $\lpf,\rpf$ of $\RPF$ and the highest eigenvalue $\spsi$ using a standard power-method series.
\item From this eigenvalue, compute the topological pressure. This gives $\tilde{h}$.
\item From the left and right eigenvectors, we have the Gibbs distribution $\mu$ corresponding to this set of parameters $\lambda_l$. One then computes the average value of $\phi_l$ under $\mu$, $\moy{\phi_l}$
Now, from (\ref{moy_phil}) the derivative of $P(\psi)$ with respect to $\lambda_l$ is exactly $\moy{\phi_l}$. This provides an exact expression for the gradient of $P(\psi)$.
\item To update the $\lambda_l$ toward the minimum of
$\tilde{h}$, we have tried several methods.  The most efficient are based on gradient algorithms where the gradient of $P(\psi)$ is exactly known from the previous step. The most efficient method seems to be the 
Fletcher-Reeves conjugate gradient algorithm from the GSL \url{http://www.gnu.org/software/gsl}, while other methods such as the Polak-Ribiere conjugate gradient algorithm, 
and the Broyden-Fletcher-Goldfarb-Shannon quasi-Newton method appeared to be less efficient. We have also used the GSL implementation of the simplex algorithm of Nelder and Mead which does not require the explicit computation of a gradient. This alternative is usually less efficient than the previous methods. All these methods are available in our library \url{http://enas.gforge.inria.fr/}.
\item Repeat the previous steps until $\tilde{h}$ attains its minimum. 
\end{enumerate}

\end{itemize}

\section{Analysis of Biological data}

\subsection{Methods}
Retinae from the larval tiger salamander (Ambystoma tigrinum) were isolated from the eye, placed over a multi-electrode array and perfused with oxygenated Ringer's medium at room temperature (22 \degree C). 
Extracellular voltages were recorded by a microelectrode array and streamed to disk for offline analysis. Spike sorting was performed as described earlier in (\cite{segev-berry:04}) to extract 40 cells.
The stimulus was a natural movie clip showing a woodland scene. The 20-30 s movie segment was repeated many times. All visual stimuli were displayed on an NEC FP1370 monitor and projected onto the retina using standard optics. The mean light level was 5 lux, corresponding to photopic vision. The total recording time was around 3200s with sampling frequency of 10000Hz.

\subsection{Analysis of spike train-statistics}
We have used the recorded spike trains of retinal ganglion cells to fit models with different sets of constraints.

The \textbf{Linear} model has a potential $\psi(\omega)=\sum_{i=1^N}  \lambda_i \, \omega_i(0)$
(thus constraints are only imposed on firing rates). The corresponding Gibbs probability is a Bernoulli  distribution where spikes are independent.
For a fixed range $R$, we call \textbf{All-R} a
potential containing all possible and non redundant monomials of range $R$.
For example, a monomial containing products of the form $\omega^k_i(n)$, $k>1$ is redundant since $\omega^k_i(n)=\omega_i(n)$. The next equation shows for clarity the potentials Linear, All-1, All-2 for a pair of neurons. Note that for a pair of neurons All-1 coincides with the usual Ising statistical model but for a triplet of neurons it contains an extra triplet synchronization term.
\begin{equation}\label{pot2}
\begin{split}
\text{linear} : \psi(\omega) &=\lambda_1\,\omega_1(0)+\lambda_2\,\omega_2(0).\\
\text{All-1 } : \psi(\omega) &=\lambda_1\,\omega_1(0)+\lambda_2\,\omega_2(0)+ \lambda_3\,\omega_1(0)\,\omega_2(0).\\
\text{All-2} : \psi(\omega) &=\lambda_1\,\omega_1(0)+\lambda_2\,\omega_2(0)+ \lambda_3\,\omega_1(0)\,\omega_2(0)\\
+&\lambda_4\,\omega_1(0)\,\omega_1(1) + \lambda_5\,\omega_2(0)\,\omega_2(1)\\
&+\lambda_6\,\omega_1(0)\,\omega_2(1) + \lambda_7\,\omega_1(1)\,\omega_2(0)\\ 
&+\lambda_8\,\omega_1(1)\,\omega_2(0) \,\omega_2(1) + \lambda_9\,\omega_1(0)\,\omega_1(1) \,\omega_2(0)\\
&+\lambda_{10}\,\omega_1(0)\,\omega_2(0) \,\omega_1(1) \,\omega_2(1).\\
\end{split}
\end{equation}

For the model estimation we bin the spike trains using bin sizes of $10$ ms (we obtain similar results with larger bin sizes). We estimate the model parameters and the Kullback-Leibler divergence between the model distribution and the empirical distribution. To have an error bar on the latter, we divide the raster in 15 equal subsets, and we randomly pick 13 subsets, on which we estimate the $d_{KL}$. This process is repeated many times (more than 100) and we estimate the error bar from the distribution of the $d_{KL}$ values obtained.

\begin{figure}[!ht]
\begin{center}
\includegraphics[height=5.5cm,width=0.9\textwidth]{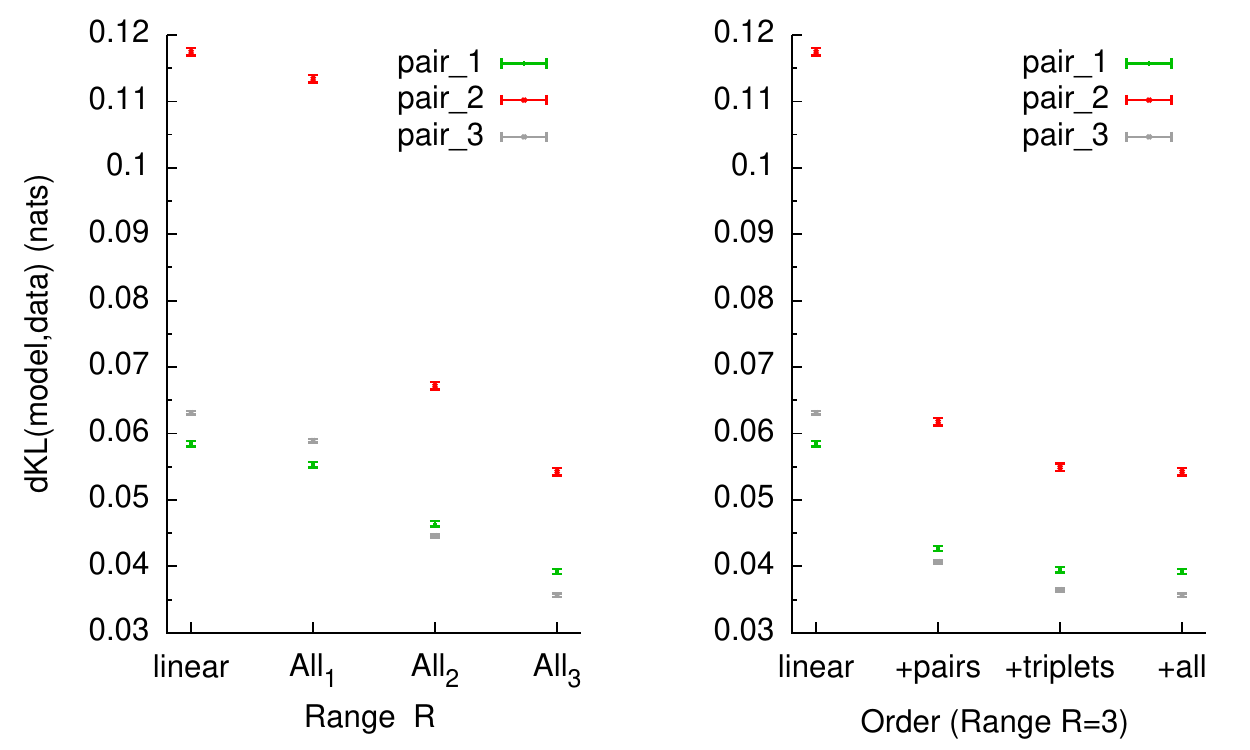} 
\end{center}
\vspace{-0.6cm}
\caption{\footnotesize{ The KL divergence between empirical distribution computed from observed data and the distribution of the estimated model (different models are shown for several pairs, and error bars are included). "nats" means "natural units" (the KL divergence is divided by $\log 2$). (Left) This figure depicts $d_{KL}$ for the Linear model and All-R from $R=1$ to $R=3$. Note that for a cell pair,  All-1 model corresponds to the pairwise Ising model. (Right) The $d_{KL}$ divergence  for models of range $R=3$ are examined versus the role of pairs, then triplets and finally the full set of terms that constitute the  All-3 model. The KL divergence of  the linear model is included  for comparison.
}}
\label{htilde}
\end{figure}

\begin{figure}[!ht]
\begin{center}
\includegraphics[height=5.5cm,width=0.9\textwidth]{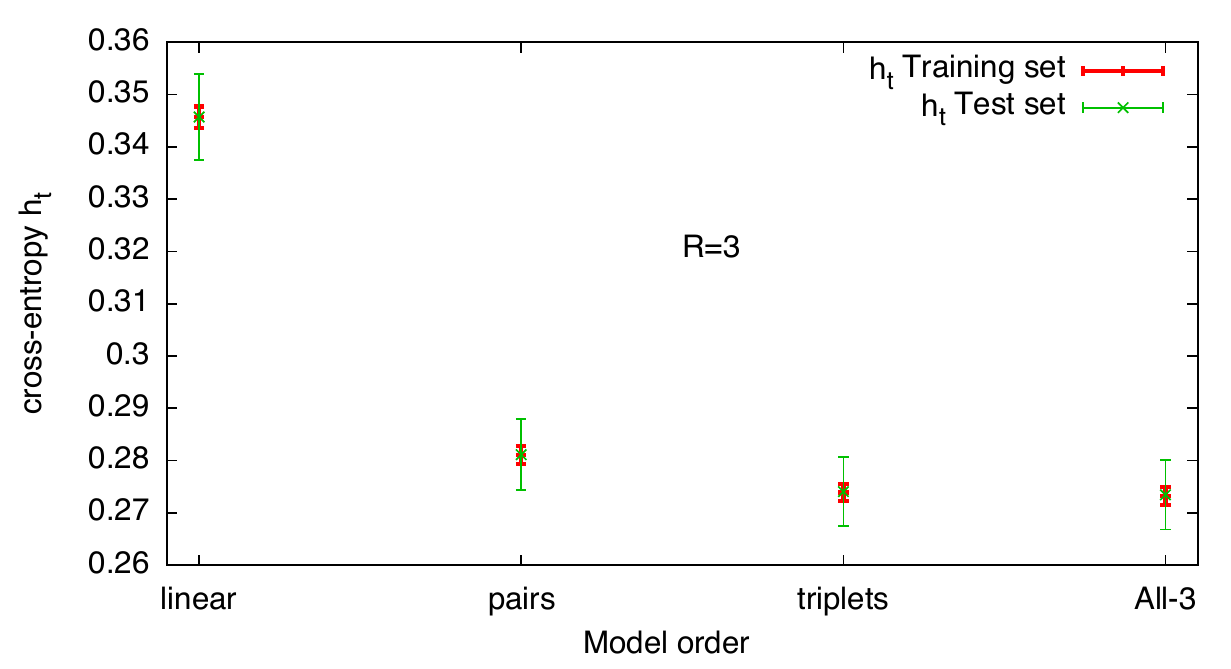} 
\end{center}
\vspace{-0.6cm}
\caption{\footnotesize{ 5-fold cross-validation for  over-fitting of the cross-entropy (written $h_t$ on the figure). 
For a single pair, the y-axis shows the cross-entropy estimated on the same models than in the previous figure, for both training and testing sets. 
}}
\label{overfitting}
\end{figure}

We first focus on the statistics of spiking patterns using models with range  from 1 up to 3. The 
KL divergence between the empirical Distribution and the model is depicted in figure \ref{htilde},  for 3 examples of pairs. Figure \ref{htilde} (left plot) shows the effect of including all interaction terms within the chosen number of time bins.  Increasing in the hierachy of models, 
from Linear to All-3 shows significant improvements.
Naturally, the number of possible interaction terms explodes combinatorially with the range of the models. Therefore, we estimate the impact of adding higher order interaction terms in a range 3 model. Figure \ref{htilde} (right plot) shows that, although the largest improvement happens when adding the pairwise terms, adding triplets interactions also allows a significant decrease of $d_{KL}$. Beyond third order, we did not see any improvement.

One could think that the improvement shown when adding monomials to the model is due to overfitting. To discard this hypothesis, we divide the raster in 5 subsets, fit the model with 4 of them, and compute the cross-entropy between the model and the fifth subset. We then change the tested subset and repeat the calculation to obtain error bars. Figure \ref{overfitting} shows that there is no difference in the mean value of the cross-entropy between the training and testing sets, and the error bar is still smaller than the difference between the models. So the improvement we see when adding terms is significant. 
Note that using the cross-entropy $\tilde{h}$, instead of Kullback-Leibler divergence $d_{KL}$, eliminates the effects of using biased  entropy estimators, as pointed out in section \ref{Scomparison}. 

\begin{figure}[!ht]
\begin{center}
\includegraphics[height=5.5cm,width=0.9\textwidth]{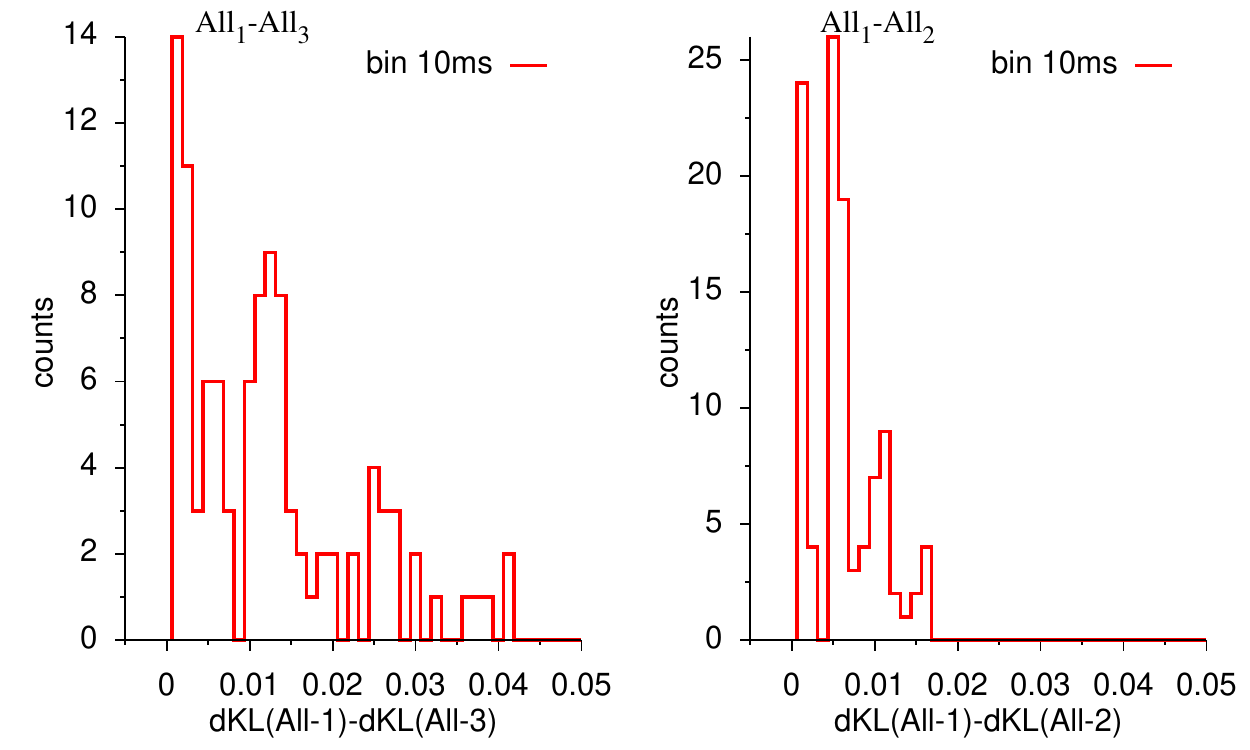} 
\end{center}
\vspace{-0.6cm}
\caption{\footnotesize{
Histogram
of differences, $\delta d_{KL}$ , between $d_{KL}$, for an All-$1$ model and an All-$3$ model (left), as well as  for an All-$1$ model and an All-$2$ model. The histogram has been computed for $100$ pairs.
}}
\label{histodeltadK}
\end{figure}

The results depicted above for the three pairs can be generalized to any pair randomly selected among the cells available. We estimate the improvement gained from the model of range 1, similar to \cite{schneidman-berry-etal:06}, or range 2, \cite{marre-boustani-etal:09}, to the model of range 3, quantified by the difference of $d_{KL}$ between the data and the range 1 or 2 model, and between the data and the range 3 model. We randomly pick $100$ pairs of cells and estimate these $d_{KL}$ differences. Fig.  \ref{histodeltadK} shows the histogram
of differences, $\delta d_{KL}$ , between $d_{KL}$ for an All-$1$ model and an All-$3$ model (left), as well as  for an All-$1$ model and an All-$2$ model  (right). Note that it is equivalent to consider $\delta \tilde{h}$ or $\delta d_{KL}$
since the term $h(\mu)$ cancels when taking the difference. The average value of $\delta d_{KL}$ for an All-$1$ model and an All-$3$ model (Fig.  \ref{histodeltadK} left) is
$0.012$ with a standard deviation $0.01$ while the average value of $\delta d_{KL}$ for an All-$1$ model and an All-$2$ model (Fig.  \ref{histodeltadK} right) is
$0.0056$ with a standard deviation $0.004$. In the former case, the difference $\delta d_{KL}$ can be more than $0.04$. So our range 3 model improves the statistical description of the data compared to previously used ones, confirming the result observed on individual pairs. The amount of improvement is highly heterogeneous depending on the pair chosen.

\begin{figure}[!ht]
\begin{center}
\includegraphics[height=5.5cm,width=0.9\textwidth]{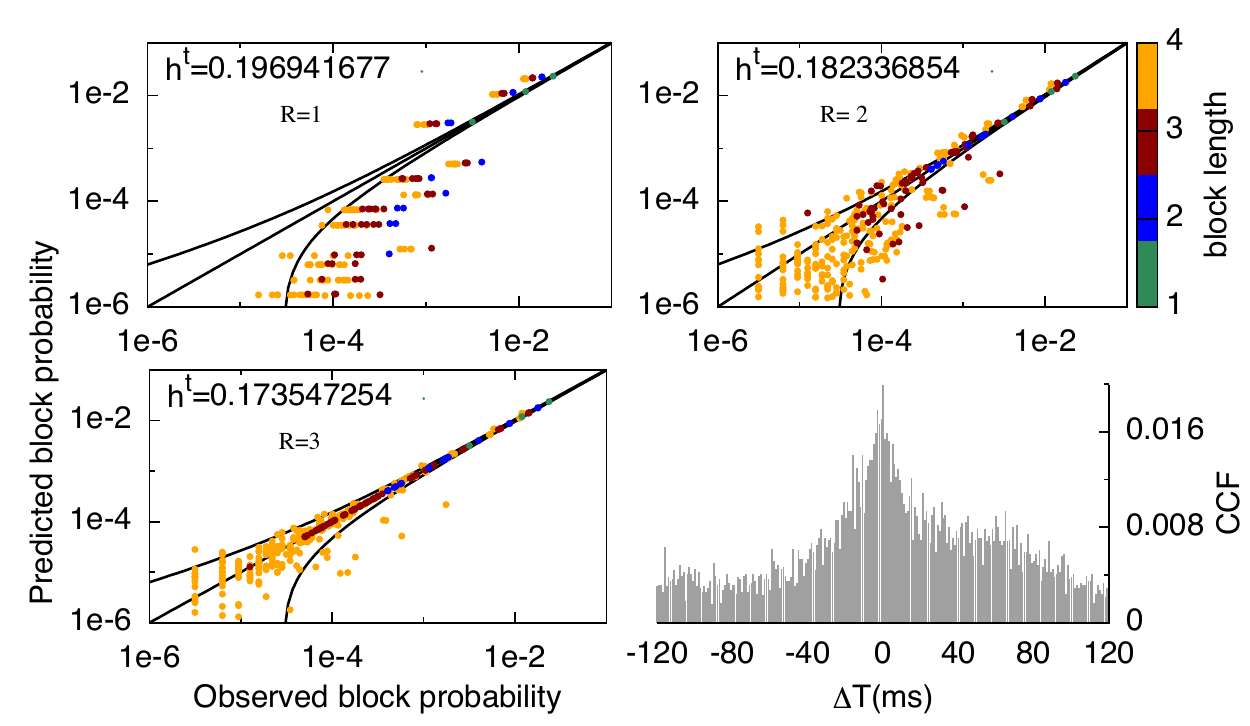} 
\end{center}
\vspace{-0.6cm}
\caption{\footnotesize{ The estimated block probability versus the observed block probability for
all possible blocks from range $1$ to $4$ (coded by colors), one pair of neurons (pair 3) using All-R models $R=1,2,3$ with data binned at 10 $ms$. We include the equality line $y=x$ and the confidence bounds (black lines) for each model, corresponding to $\mu(w)=\pT{w}{\omega} \pm 3\sigma_{w} $, $\sigma_{w}$ being the standard deviation for each estimated probability given the total sample length $T\sim 3\cdot 10^5$. The cross-correlation function (CCF) of this pair is also depicted  (bottom right).
}}
\label{p3}
\end{figure}

\begin{figure}[!ht]
\begin{center}
\includegraphics[height=5.5cm,width=0.9\textwidth]{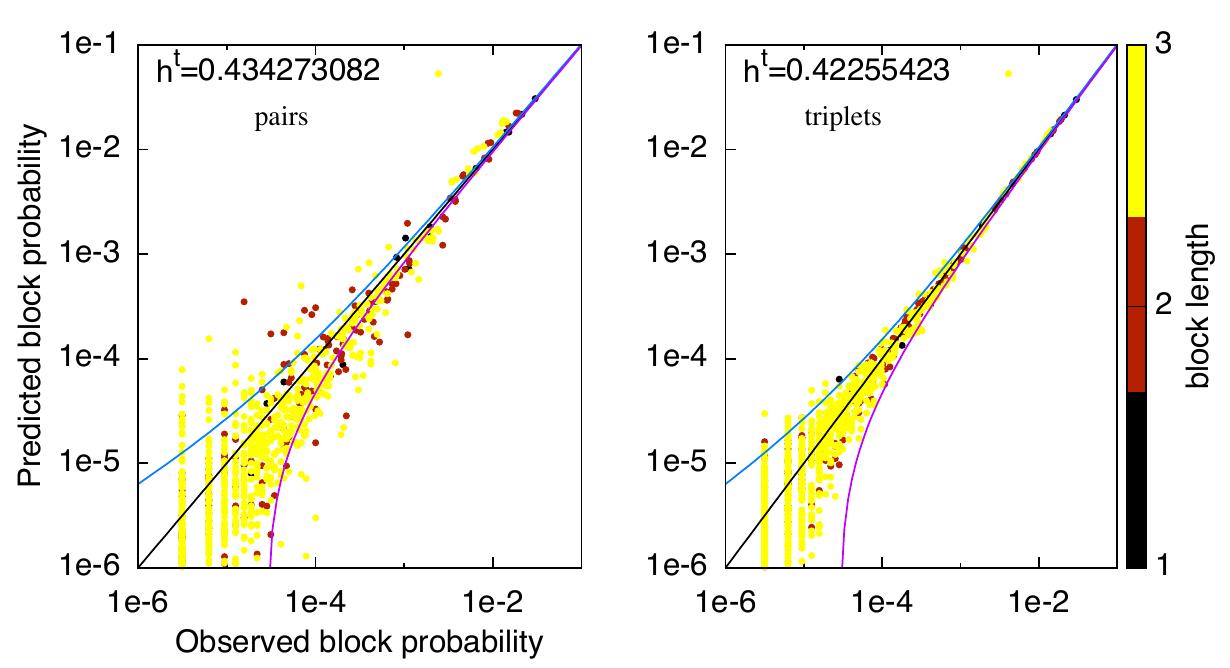} 
\end{center}
\vspace{-0.6cm}
\caption{\footnotesize{ The estimated block probability versus the observed block probability for
all blocks from range $1$ to $4$ (coded by colors), for $N=4$ neurons with a model of range $R=3$
for pairs and triplets.  Data  is binned at 10 $ms$.
}}
\label{n4p0}
\end{figure}

Can these models predict statistics on which they were not fitted? To answer that question, we estimate the rate of spiking pattern of two neurons and 1 to 4 time bins. Figure \ref{p3} shows the empirically-observed pattern rate against the pattern rate predicted by each model All-1, All-2 and All-3. Each point corresponds to a spike block. It appears that the model All-3 provides a much better description of the statistics than All-1 and All-2. The result also holds for triplets of neurons (data not shown) and still holds for a bin size of 20ms. In addition, to explore the effects of including higher order spatio-temporal interactions given a  range, we show in Figure \ref{n4p0} the same type of plot, for a set of N=4 neurons with models of range R=3 with pairs and triplets. So triplet terms do enhance statistical description of spatio-temporal patterns.

\begin{figure}[!ht]
\begin{center}
\includegraphics[height=5.5cm,width=0.9\textwidth]{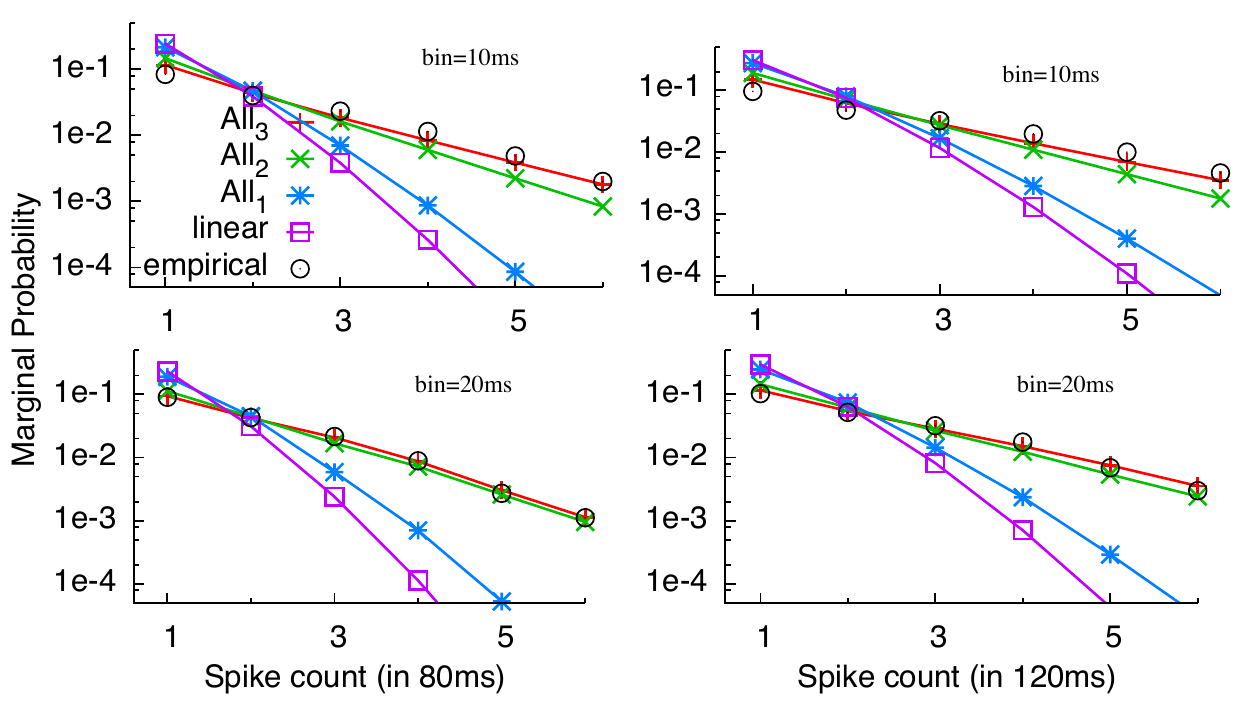} 
\end{center}
\vspace{-0.6cm}
\caption{\footnotesize{ Distribution of the number of spikes fired by a pair of cells 
in $80 ms$ (left column) and $120 ms$ (right column), compared with predictions by several models:
Linear (independent), All-1, All-2,All-3 and bin sizes $10 ms$ (upper row) and $20 ms$ (bottom row).}}
\label{spikecountmarginal}
\end{figure}

We also assess the performance of the models in predicting the total number of spikes during a given window of time. Figure \ref{spikecountmarginal} shows this
performance for several models, fitted with two different bin sizes. The number of spikes, measured or predicted, is counted over $80$ and $120$ ms windows.  The All-2 model already predicts well the statistics, and the All-3 model improves marginally the performance.  These two models are visually almost indistinguishable  when a small number of bins is used  (i.e, bin size of 20ms corresponding to bottom row uses 4 and 6 bins, while the bin size of 10ms depicted on the upper row uses 8 and 12 bins).

\begin{figure}[!ht]
\begin{center}
\includegraphics[height=5.5cm,width=0.9\textwidth]{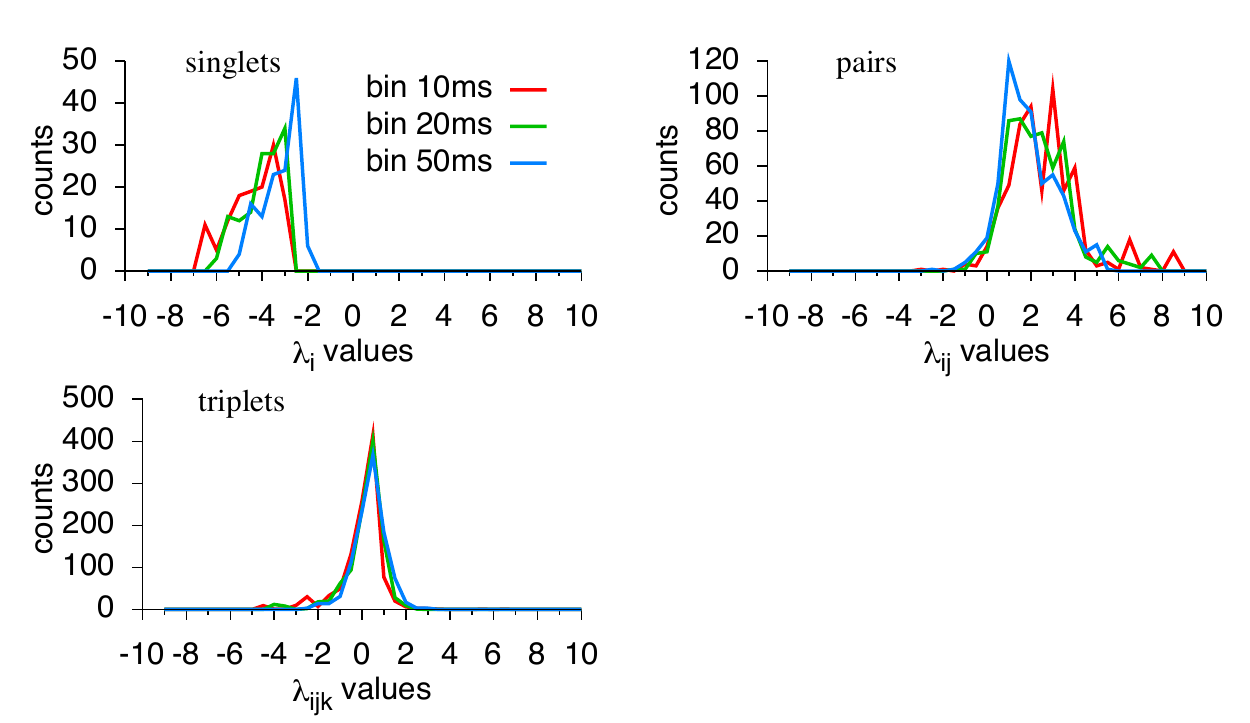} 
\end{center}
\vspace{-0.6cm}
\caption{\footnotesize{ Distribution of the monomial coefficients values has been computed  after estimation of a All-3 model over 50 different pairs, for several choices of bin size. The histogram has been constructed separately for single spikes, pairs and triplets. Note that in our framework we consider only non redundant monomials so there is a single coefficient for each monomial. For instance, for the monomial $\omega_i(0)\,\omega_j(0)$ there is only  one coefficient $ij$ without a symmetrical $ji$ present 
}}
\label{histolambda}
\end{figure}

We examine the coefficients of the parametric estimation by plotting the distribution of the
monomial coefficients values after estimation of a All-3 model over 50 different pairs
for single spikes, pairs, and triplets. They are depicted  in figure \ref{histolambda}. Note that none of them is centred at zero,  in particular triplet terms are not negligible, suggesting that higher order spatio-temporal interactions do matter. The same conclusion  holds for groups of three neurons (data not shown). Additionally, we remark that taking larger bin sizes ($20,50$ms)  reduces the relative value of coefficients but distribution is still not centred.

\section{Discussion and Conclusion}
\label{Sec:discussion}

In this paper, we have developed a Gibbs distribution analysis for general spatio-temporal spike patterns. Our method allows one to handle Markovian models with memory up to the limits imposed by the finite size of the data. Our analysis on retina data 
suggests that higher order interaction terms, as well as interaction between non consecutive time bins, are necessary to model the statistics of the spatio-temporal spiking patterns, at least for small populations of neurons.  

An important issue is to determine whether these higher order terms are still essential when looking at much larger groups of neurons: either the complexity of the models will grow with the number of neurons, or adding neurons will have a similar effect as uncovering hidden variables, and might then weaken these interactions. The extension to large networks of our method is, thus, an important future step to progress in our understanding of the spatio-temporal statistics of spike trains. 
However, the identification of the relevant neural subsets in a large number of neurons remains an open problem. 
Moreover, to explore models with larger ranges, one needs to control the confidence level due to finite size effects given the available amount of data, which can be addressed through Neyman-Pearson results for Markov chains of finite order \cite{nagaev:02}.  Additionally, the spectrum of the Perron-Frobenius matrix provides information about the correlation decay time, which can be used to determine the optimal range of the model. Both issues are to be developed in a forthcoming paper. 

\vspace{0.2in}

\small{
\textbf{Acknowledgments}
This work has highly benefited from the collaboration with the INRIA team Cortex, and especially T. Vi\'eville. It
was supported by the INRIA, ERC-NERVI number 227747, KEOPS ANR-CONICYT and European Union Project FP7-269921 (BrainScales) projects to B.C and J.C.V; grants EY 014196 and EY 017934 to M.J.B; and FONDECYT 1110292, ICM-IC09-022-P  to A.P.

 J.C Vasquez has been funded by French ministry of Research and University of Nice (EDSTIC). 
}

\bibliographystyle{elsarticle-num-names}
{\scriptsize \bibliography{biblio-gibbsgramme,biblio,odyssee}}

\end{document}